%% file: fhe-accelerator-fpga-2022.tex
\documentclass{sig-alternate}
\usepackage{mathptmx} 

\usepackage{fancyhdr}
\usepackage[normalem]{ulem}
\usepackage[hyphens]{url}
\usepackage[sort,nocompress]{cite}
\usepackage[final]{microtype}
\usepackage[keeplastbox]{flushend}
\usepackage{enumitem}

\usepackage{breakcites}

\usepackage{amsmath,amssymb,amsfonts}

\usepackage{algorithm}
\usepackage{graphicx}
\usepackage{textcomp}

\usepackage[toc,page]{appendix}

\usepackage{soul}

\usepackage{color}
\usepackage[usenames,dvipsnames]{xcolor}

\usepackage{xspace}
\usepackage{threeparttable, booktabs, tabularx}
\usepackage[hyphens]{url}
\usepackage{stmaryrd}
\usepackage{hyperref}
\hypersetup{
    colorlinks=true,
    linkcolor=magenta,
    filecolor=magenta,
    citecolor=red,
    urlcolor=cyan,
}
\usepackage[hyperpageref]{backref}
\usepackage{cleveref}
\usepackage{algpseudocode}
\usepackage{tablefootnote}

\usepackage{multirow}

\newcommand{\ignore}[1]{}
\newcommand{\fab}{FAB\xspace}

\def\BibTeX{{\rm B\kern-.05em{\sc i\kern-.025em b}\kern-.08em
    T\kern-.1667em\lower.7ex\hbox{E}\kern-.125emX}}

\input{Preamble}

\title{FAB: An FPGA-based Accelerator for\\ Bootstrappable Fully Homomorphic Encryption} 

\author{
    Rashmi Agrawal$^1$
	Leo de Castro$^2$
	Guowei Yang$^1$
	Chiraag Juvekar$^3$
	Rabia Yazicigil$^1$ \\
	Anantha Chandrakasan$^2$
	Vinod Vaikuntanathan$^2$
	Ajay Joshi$^1$\\
	\small $^1$Boston University, Boston MA, USA;
	\small $^2$MIT, Cambridge, MA, USA;
	\small $^3$Analog Devices, Boston, MA USA\\
	\small \{rashmi23, guoweiy, rty, joshi\}@bu.edu, \{ldec, anantha, vinodv\}@mit.edu, chiraag.juvekar@analog.com
}

\begin{document}
\maketitle

\pagestyle{plain}

\begin{abstract}
\vspace{-0.2in}
Fully Homomorphic Encryption (FHE) offers protection to private data on third-party cloud servers by allowing computations on the data in encrypted form.
However, to support general-purpose encrypted computations, all existing FHE schemes require an expensive operation known as ``bootstrapping''.
Unfortunately, the computation cost and the memory bandwidth required for bootstrapping add significant overhead to FHE-based computations, limiting the practical use of FHE.

In this work, we propose \fab, an FPGA-based accelerator for bootstrappable FHE.
Prior FPGA-based FHE accelerators have proposed hardware acceleration of basic FHE primitives for impractical parameter sets without support for bootstrapping.
\fab, for the first time ever, accelerates bootstrapping (along with basic FHE primitives) on an FPGA for a secure and practical parameter set.
Prior hardware implementations of FHE that included bootstrapping are heavily memory bound, leading to large execution times and wasted compute resources.
The key contribution of our work is to architect a balanced \fab design, which is not memory bound. 
To this end, we leverage recent algorithms for bootstrapping while being cognizant of the compute and memory constraints of our FPGA.
To architect a balanced \fab design, we use a minimal number of functional units for computing, operate at a low frequency, leverage high data rates to and from main memory, utilize the limited on-chip memory effectively, and perform operation scheduling carefully.

We evaluate \fab using a single Xilinx Alveo U$280$ FPGA and by scaling it to a multi-FPGA system consisting of eight such FPGAs.
For bootstrapping a fully-packed ciphertext, while operating at $300$~MHz, \fab outperforms existing state-of-the-art CPU and GPU implementations by $213\times$ and $1.5\times$ respectively.
Our target FHE application is training a logistic regression model over encrypted data.
For logistic regression model training scaled to $8$~FPGAs on the cloud, \fab outperforms a CPU and GPU by $456\times$ and $6.5\times$, and provides competitive performance when compared to the state-of-the-art ASIC design at a fraction of the cost. 
\end{abstract}

\input{Sections/Introduction}
\input{Sections/Background}
\input{Sections/FABArchitecture}
\input{Sections/FABMicroarchitecture}
\input{Sections/Evaluation}
\input{Sections/RelatedWork}
\input{Sections/Conclusion}

\bibliographystyle{IEEEtranS}
\bibliography{fhe-accelerator-fpga-2022}

\end{document}

%% file: Preamble.tex
\newcommand{\ring}{\mathcal{R}}

\newcommand{\C}{\mathbb{C}}

\newcommand{\Z}{\mathbb{Z}}

\newcommand{\Add}{\mathsf{Add}}

\newcommand{\Automorph}{\mathsf{Automorph}}
\newcommand{\ModUp}{\mathsf{ModUp}}

\newcommand{\ModDown}{\mathsf{ModDown}}

\newcommand{\Mult}{\mathsf{Mult}}

\newcommand{\KeySwitch}{\mathsf{KeySwitch}}
\newcommand{\Rotate}{\mathsf{Rotate}}

\newcommand{\Conjugate}{\mathsf{Conjugate}}

\newcommand{\Rescale}{\mathsf{Rescale}}
\newcommand{\KSKInProd}{\mathsf{KSKIP}}

\newcommand{\calB}{\mathcal{B}}

\newcommand{\BasisConvert}{\mathsf{BasisConvert}}

\newcommand{\NTT}{\mathsf{NTT}}

\newcommand{\ksk}{\mathsf{ksk}}

\newcommand{\fftIter}{\mathsf{fftIter}}

\newcommand{\dnum}{\mathsf{dnum}}

\newcommand{\Decomp}{\mathsf{Decomp}}

\newcommand{\dbrack}[1]{\left\llbracket #1 \right\rrbracket}

\renewcommand{\a}{\mathbf{a}}
\renewcommand{\b}{\mathbf{b}}
\renewcommand{\c}{\mathbf{c}}

\newcommand{\m}{\mathbf{m}}
\newcommand{\s}{\mathbf{s}}

\newcommand{\x}{\mathbf{x}}

%% file: Sections/Introduction.tex
\section{Introduction}
\label{sec:Introduction}
The last decade has seen rapid growth in machine learning (ML) with exciting ML-based applications in a variety of fields.
Given the sheer amount of data that is used by these applications, ML accelerators on the cloud~\cite{zhang2019mark,kachris2016survey} offer a way to accelerate the performance of each phase (data storage, training, and inference) in ML.
However, these cloud-based accelerators also introduce privacy concerns by providing unrestricted access to the data to third-party cloud servers.

Fully homomorphic encryption (FHE)~\cite{RAD,Gentry09} is currently regarded as the ``gold standard'' to preserve the privacy of data while enabling machine learning training/inference in an untrusted cloud environment. 
FHE enables computing on data \emph{while it is encrypted}, allowing the cloud to operate on the data without having access to the data itself.
Despite this incredible achievement of theoretical cryptography, the large compute and memory requirements of FHE remain a serious barrier to its widespread adoption.
For example, to train a logistic regression (LR) model for $30$~iterations using plaintext data ($11,982$ samples with $196$ features), it takes ${\sim}1.05$~sec on a CPU.
The same LR model training on encrypted data takes ${\sim}124$~mins on the same CPU~\cite{han2018efficient}, a slowdown of about $7086\times$.
At the same time, the size of encrypted data ($198$~MB) used in this training is $152\times$ larger than the size of the plaintext data ($1.3$~MB).

To address the compute and memory requirements of FHE, several optimizations and acceleration efforts are in progress. 
For FHE-based computing on CPUs, SEAL~\cite{sealcrypto}, PALISADE \cite{PL2019}, HELib~\cite{HElib,HS2014}, NFLLib~\cite{AN2016}, Lattigo~\cite{LT2019}, and HEAAN \cite{KH2018} software libraries accelerate one or more FHE schemes.
Unfortunately, CPUs do not have the capability to adequately exploit the inherent parallelism available in FHE. 
GPU-based FHE accelerators~\cite{dai2015cuhe,al2018high,jung2021over} have recently had more success.
These GPU-based implementations exploit the inherent parallelism in FHE, but the GPUs have massive floating-point units that are completely underutilized as FHE-based workloads consist almost entirely of integer-only operations. 
Moreover, neither CPU nor GPU can provide adequate main memory bandwidth to handle the data-intensive nature of the FHE-based workloads.

Consequently, Samardzic et al.~\cite{samardzic2021f1} and Kim et al.~\cite{kim2021bts} proposed custom FHE accelerators called F$1$ and BTS, respectively.
Both accelerators are highly resource-intensive, requiring a massive number ($8192$-$18432$) of processing elements and large on-chip memory ($64$-$512$~MB) to accelerate FHE workloads. 
Despite these resources, neither F$1$ nor BTS is able to store the entire working set of FHE-based workloads. 
Effectively, these proposals are still bottlenecked by main memory bandwidth. 
Since main memory cannot keep up with the custom ASIC implementations of F$1$ and BTS, this leads to the compute portions of these accelerators sitting idle. 
A more in-depth discussion of these related works is in Section~\ref{sec:RelatedWork}. 
At a high-level, neither of these ASIC proposals overcome the fundamental barrier of  memory bandwidth, which limits their performance.

In this work, we propose \fab, an FPGA-based accelerator for bootstrappable FHE that supports the Cheon-Kim-Kim-Song (CKKS)~\cite{CKKS17} FHE scheme. 
\fab makes use of state-of-the-art analysis of the bootstrapping algorithm~\cite{fhecompute2021} to co-design the FHE operations and select parameters that are optimized for the hardware constraints. 
This allows \fab to support practical FHE parameter sets (i.e. parameters large enough to support bootstrapping) \emph{without} being bottlenecked by the main-memory bandwidth.
This does not come at the cost of compute efficiency. 
We evaluate \fab on a target application of training a logistic regression model over encrypted data. 
Our benchmarks demonstrate that \fab outperforms all prior works ($6.5\times$ to $456\times$) and is competitive with the state-of-the-art ASIC proposal. 
More details on our evaluation are in Section~\ref{sec:Evaluation}. 

We architect \fab for the Xilinx Alveo U$280$ FPGA accelerator card containing High Bandwidth Memory 2 (HBM$2$).
\fab is highly resource efficient, implementing only $256$ functional units, where each functional unit supports various modular arithmetic operations.
\fab exploits maximal pipelining and parallelism by utilizing these functional units as per the computational demands of the FHE operations.   
\fab also makes efficient use of the limited $43$~MB on-chip memory and a $2$~MB register file to manage the ${>}100$~MB working dataset at any given point of time. 
Moreover, \fab leverages a smart operation scheduling to enable higher data reuse and prefetching of the required datasets from main memory without stalling the functional units.
In addition, this smart scheduling evenly distributes the accesses to main memory so as to efficiently utilize the limited main memory bandwidth through a homogeneous memory traffic. A detailed description of the microarchitecture of \fab is given in Section~\ref{sec:Microarchitecture}. 

The performance of \fab suggests that FPGA is a ``sweet-spot'' for FHE acceleration. \fab significantly outperforms both CPU and GPU implementations of FHE. In contrast to ASICs, \fab only uses standard, commercially available hardware that is highly accessible to the general public (e.g. via the AWS F$1$ cloud). By enabling near-ASIC levels of performance with the same availability as a high-end GPU, \fab demonstrates that FPGAs are the most viable option for near-term hardware acceleration of FHE. Furthermore, \fab effectively raises the bar for the performance of future ASIC proposals since the advantage of the current state-of-the-art ASIC proposal over \fab is not worth the millions of dollars of investment required to fabricate a custom hardware chip.

In summary, we make the following contributions:
\begin{itemize}[leftmargin=*]
    \item We propose \fab, a novel accelerator that supports all homomorphic operations, including fully-packed bootstrapping, in the CKKS FHE scheme for practical FHE parameters. This accelerator outperforms all prior works and is competitive with the latest ASIC proposals for our target application: secure training of logistic regression models. 
    
    \item \fab tackles the memory-bounded nature of bootstrappable FHE through smart operation scheduling and on-chip memory management techniques, in turn maximizing the overall FHE-based computing throughput. 
    
    \item Last but not least, \fab uses only currently-existing hardware (FPGAs), and {\em does not require any custom hardware}. This makes \fab immediately accessible to the general public as all of the resources required to support \fab exist in public commercial cloud environments (e.g. AWS F$1$ instances). In short, \fab offers performance that is competitive with ASIC with the accessibility of a pure CPU/GPU implementation, and demonstrates FPGAs to be a sweet-spot for bootstrappable FHE.
\end{itemize}

%% file: Sections/Background.tex
\section{Background}
\label{sec:Background}
In this section, we briefly review the CKKS~\cite{CKKS17} homomorphic encryption scheme and the relevant parameters for FAB. A summary of these parameters is given in Table~\ref{tab:parameters}.

\begin{table}[t]
    \centering
    \caption{CKKS FHE Parameters and their description.}
    \label{tab:parameters}
    \begin{tabular}{p{0.2\columnwidth} p{0.7\columnwidth}}
    \toprule
    \textbf{Parameter} & \textbf{Description}\\
    \midrule
    $N$ & Number of coefficients in the ciphertext polynomial.\\
    $n$ & Number of plaintext elements in a ciphertext ($n \leq N/2$ is required).\\
    $Q$ & Full modulus of a ciphertext coefficient.\\
    $q$ & Prime modulus and a limb of $Q$.\\
    $L$ & Maximum number of limbs in a ciphertext.\\
    $\ell$ & Current number of limbs in a ciphertext.\\
    $\dnum$ & Number of digits in the switching key.\\
    $\alpha$ & $\lceil (L+1)/\dnum \rceil$. Number of limbs that comprise a single digit in the key-switching decomposition. This value is fixed throughout the computation. \\
    $P$ & Product of the extension limbs added for the raised modulus. There are $\alpha+1$ extension limbs.\\
    $\fftIter$ & Multiplicative depth of a linear transform in bootstrapping.\\
    \bottomrule
    \end{tabular}
   % \vspace{-0.2in}
\end{table}

\subsection{The CKKS FHE Scheme}
\label{subsec:CKKS}
The CKKS~\cite{CKKS17} scheme supports operations over vectors of complex numbers.
A plaintext in the CKKS scheme is an element of $\C^n$, where $\C$ is the field of complex numbers. 
The plaintext operations are component-wise addition and component-wise multiplication of elements of $\C^n$. 
In addition to the plaintext size $n$, CKKS is parameterized by a ciphertext coefficient modulus $Q \in \Z$ (where $\Z$ is the ring of integers) and a ciphertext polynomial modulus $x^N+1$, where $N$ is chosen to be a power of $2$. 
CKKS ciphertexts are elements of $\ring_Q^2$, where $\ring_Q := \Z_Q[x]/(x^N+1)$.
We denote the encryption of a vector $\m \in \C^n$ by $\dbrack{\m} = (\a_\m, \b_\m)$ where $\a_\m$ and $\b_\m$ are the two elements of $\ring_Q$ that comprise the ciphertext.

CKKS supports the following operations over encrypted vectors. All arithmetic operations between two plaintext vectors are \emph{component-wise}.
\begin{itemize}
    \item $\Add(\dbrack{\m_1}, \dbrack{\m_2}) \rightarrow \dbrack{\m_1 + \m_2}$, where the addition is component-wise over $\C$.
    \item $\Mult(\dbrack{\m_1}, \dbrack{\m_2}) \rightarrow \dbrack{\m_1 \odot \m_2}$, where $\odot$ represents the component-wise product of two vectors.
    \item $\Rotate(\dbrack{\m}, k) \rightarrow \dbrack{\phi_k(\m)}$, where $\phi_k$ is the function that rotates a vector by $k$ entries.
     As an example, when $k = 1$, the rotation $\phi_1(\x)$ is defined as follows: 
    \begin{align*}
        \x &= \begin{pmatrix}
        x_0 & x_1 & \ldots & x_{n-2} & x_{n-1}
        \end{pmatrix}\\
        \phi_1(\x) &=\begin{pmatrix}
        x_{n-1} & x_{0} & \ldots & x_{n-3} & x_{n-2}
        \end{pmatrix}
    \end{align*}
    \item $\Conjugate(\dbrack{\m}) \rightarrow \dbrack{\overline{\m}}$ where $\overline{\ \cdot\ }$ represents the complex conjugate operation.
\end{itemize}

\subsubsection{Homomorphic Levels \& RNS Representation} 
\label{subsec:RNS}
To efficiently operate over elements of $\ring_Q$, we represent $Q$ as a product of primes $q_1, \ldots, q_\ell$ where each $q_i$ is roughly the size of a machine word. This follows the standard residue number system (RNS) representation of ciphertext moduli~\cite{BFVRNS, FullRNSHEAAN}. We call each $q_i$ a \emph{limb} of the modulus $Q$, and we say a modulus $Q := \prod_{i=1}^\ell q_i$ has $\ell-1$ \emph{levels}. 
We call the set $\calB := \{q_1, \ldots, q_\ell\}$ an \emph{RNS basis}. 
Each multiplication operation reduces the size of the modulus by one limb. A modulus with $\ell$ levels can support a circuit of multiplicative depth $\ell$ before bootstrapping is required. Addition, rotation, and conjugation operations do not change the number of levels in a modulus. 

This representation allows us to operate over values in $\Z_Q$ without any native support for multi-precision arithmetic. 
Instead, we can represent $x \in \Z_Q$ as a length-$\ell$ vector of scalars $[x]_\calB = (x_1, x_2, \ldots, x_\ell)$, where $x_i \equiv x \pmod{q_i}$. 
We refer to each $x_i$ as a \emph{limb} of $x$.
To add two values $x, y \in \Z_Q$, we have $x_i + y_i  \equiv x + y \pmod{q_i}$. Similarly, we have $x_i \cdot y_i  \equiv x \cdot y \pmod{q_i}$. 
This allows us to compute addition and multiplication over $\Z_Q$ while only operating over standard machine words. 

We briefly define the equation for RNS recombination. We refer the reader to \cite{BFVRNS, FullRNSHEAAN} for more details. This equation takes in an RNS representation of a value $x \in \Z_Q$ as a length-$\ell$ vector of scalars $[x]_\calB = (x_1, x_2, \ldots, x_\ell)$, where $x_i \equiv x \pmod{q_i}$. It outputs $x \mod p$, where $p$ is a new modulus in an extended RNS basis.
\begin{align}\label{eq:RNSExtension} 
    [x]_p = \sum_{i=1}^\ell [x_i \cdot \tilde{Q}_i]_{q_i} \cdot Q_i^* \pmod p 
\end{align}
where $Q_i^* = Q/q_i$ and $\tilde{Q}_i = (Q_i^*)^{-1} \pmod q_i$

In memory, ciphertext data can be viewed as an $\ell \times n$ matrix, where each row is a limb and each column corresponds to a single coefficient modulo $Q$. Arranging this matrix in "row-major order" where there is locality for elements in the same row is called \emph{limb-wise} access, since it enables efficient data within a single limb. By contrast, arranging this matrix in "column-major order" where there is locality for elements in the same column is known as \emph{slot-wise} access, since it is best for accessing the same slot of data across all of the ciphertext limbs.

\subsubsection{Number Theoretic Transform} 
\label{subsec:NTT_BG}
To efficiently multiply elements of $\ring_Q$, we make use of the number theoretic transform (NTT), which is the analog of the Fast Fourier Transform (FFT) modulo $q$. All polynomials in the CKKS scheme are represented by default as a series of $N$ evaluations at fixed roots of unity,  allowing fast polynomial multiplications (in $O(N)$ time instead of $O(N^2)$). 
NTT is the finite field version of the fast Fourier transform (FFT) and takes $O(N\log N)$ time and $O(N)$ space for a degree-$(N-1)$ polynomial.
We call the output of the NTT on a polynomial its \emph{evaluation representation}. 
If any operations are to be done over the polynomial's \emph{coefficient representation}, then we need to perform an inverse NTT (iNTT) to move the polynomial back to its \emph{coefficient representation}.

We would like to note that in addition to NTT polynomial transform the CKKS scheme also uses the FFT polynomial transform. 
On the client side, during CKKS encryption and decryption, a complex FFT must be run on vectors of complex numbers to map them to polynomials that can be encrypted. 
Correspondingly, on the cloud side, during bootstrapping, this complex FFT (which was performed on the client side) must be \emph{homomorphically} evaluated on the encrypted data.

\subsubsection{Bootstrapping}
In order to compute indefinitely on a CKKS ciphertext, there needs to be an operation that raises the ciphertext modulus $Q$ while maintaining the correct structure of the ciphertext. This operation is known as bootstrapping~\cite{CKKS20}, and it is the main bottleneck for fully-homomorphic encryption. Optimizing bootstrapping is the main focus of \fab. 

A detailed description of the bootstrapping algorithm is beyond the scope of this work. We refer the reader to \cite{fhecompute2021} for a detailed description of the bootstrapping algorithm that we use. This analysis is particularly relevant for \fab because it optimizes for the memory bandwidth of the bootstrapping algorithm. Without this analysis and the resulting optimizations, \fab would also be bottlenecked by the memory bandwidth; however, with these optimizations, the memory bandwidth is no longer the bottleneck. 

At a high level, the bootstrapping operation consists of three major steps: a linear transform, a polynomial evaluation, and finally another linear transform (inverse of the first step). All these steps consist of the same homomorphic operations described above ($\Add, \Mult, \Rotate,$ and $\Conjugate$).

The polynomial evaluation is constrained by application parameters, and we use the same polynomial as Bossuat et al.~\cite{BMTH20} to support non-sparse CKKS secret keys. The multiplicative depth of this polynomial evaluation is 9. 

The two linear transforms in bootstrapping are FFT and inverse FFT, which must be homomorphically evaluated on the encrypted data. There is a depth-performance trade-off in this algorithm that has been carefully studied in prior works~\cite{CCS18, CHH18, fhecompute2021}. This trade-off is parametrized by the chosen multiplicative depth of the FFT algorithm, which we denote as $\fftIter$. A more detailed discussion of the effects of this parameter is given Section~\ref{subsec:Parameter}. 

\subsubsection{Bootstrapping Performance} We review the main performance metric that is used to evaluate the efficiency of bootstrapping, first introduced in \cite{jung2021over}. After a bootstrapping operation, the resulting ciphertext can support a certain number of computation levels $\ell$ before needing to be bootstrapped again. Since each level corresponds to a multiplication, the performance metric for a bootstrapping routine is as follows:
\begin{align} \label{eq:AmortizedMultTime}
    T_{\Mult, a/\mathsf{slot}} := \frac{T_{\mathsf{Boot}} + \sum_{i=1}^\ell T_{\Mult}(i)}{\ell \cdot n}
\end{align}
This is known as the \emph{amortized multiplication time per slot}.
Here, $n$ is the number of slots in the ciphertext, $T_{\mathsf{Boot}}$ is the bootstrapping time, and $T_{\Mult}(i)$ is the time to multiply at level $i$. 

The number of levels $\ell$ in the resulting ciphertext are equal to the maximum supported levels in the starting bootstrapping modulus minus the depth of bootstrapping. The depth of bootstrapping for our algorithm is $L_{\mathsf{Boot}} := 2\cdot \fftIter + 9$.

\subsubsection{Switching Keys \& Major Subroutines}
The $\Mult$, $\Rotate,$ and $\Conjugate$ operations all produce intermediate ciphertexts that are decryptable under a different secret key than the input ciphertext. These operations all use a common subroutine known as $\KeySwitch$~\cite{BV11} to switch the decryption key back to the original value. $\KeySwitch$ represents a major bottleneck in low-level homomorphic operations.

All instances of $\KeySwitch$ require a \emph{switching key}, which is a special type of key generated using the secret key but safe to publish along with the rest of the public key. The $\KeySwitch$ operation takes in a switching key $\ksk_{\s \rightarrow \s'}$ and a ciphertext $\dbrack{\m}_{\s}$ decryptable under a secret key $\s$ and produces a ciphertext $\dbrack{\m}_{\s'}$ that encrypts the same message but can be decrypted under a different key $\s'$. We use the structure of the switching key proposed by Han and Ki~\cite{HK19}, where the switching key is parameterized by a length $\dnum$ and is a $2 \times \dnum$ matrix of polynomials.

\begin{align} 
\label{eq:kskShape}
\ksk = \begin{pmatrix} 
\a_1 & \a_2 & \ldots & \a_\dnum \\ 
\b_1 & \b_2 & \ldots & \b_\dnum \\ 
\end{pmatrix}
\end{align} 

We omit the descriptions of the CKKS subroutines. 
We use the same definitions as \cite{fhecompute2021}, and we refer the reader to \cite{fhecompute2021} for a thorough description of these subroutines. 
For reference, we make use of the $\Decomp$, $\ModUp$, $\ModDown$, $\mathsf{KeySwitchInnerProd}$ (which we call $\KSKInProd$), and $\Automorph$ subroutines.

During the course of key switching, the ciphertext coefficient modulus is raised from $Q$ to $P\cdot Q$, where $P$ is a fixed product of "extension limbs." The ciphertext modulus $Q$ is split into at most $\dnum$ digits of equal size, and $P$ must be larger than the largest product of the limbs in a single digit of $Q$. We refer the reader to \cite{HK19} for more details, and we conclude by noting that the parameter $P\cdot Q$ is the maximum modulus for which security must be maintained.  

\subsection{Practical Parameter Set for FAB}
\label{subsec:Parameter}
To prototype an efficient \fab design on the Xilinx Alveo U$280$ FPGA, we identify an optimal FHE parameter set that can support CKKS bootstrapping as well as the computational requirement of a real-time machine learning application. 
The parameter $N$ must be a power of two for efficiency of the NTT, and the largest power of two that still leaves the limbs small enough to fit in the on-chip memory is $N = 2^{16}$. Given this $N$, 
the maximum ciphertext modulus we can support is $\log(PQ)=1728$, which achieves a $128$-bit security level~\cite{lweEstimator, BMTH20}. These parameters also meet the constraints of our chip, since the maximum size of a single ciphertext is only $28.3$~MB (based on the maximum number of raised limbs, which is $32$ limbs). We can fit an entire ciphertext in the limited on-chip memory ($43$~MB) of our FPGA and thus limit the data movement to \& from the main memory.
Table~\ref{tab:parameterset} lists our choice of the other parameters based on the selected values for $\log(PQ)$ and $N$. 

\textbf{Limb Bit-Width.}
We fix the bit-width for each limb ($\log q$) as $54$ bits for several reasons.
First, a $54$-bit limb width enables effective utilization of both the $18$-bit multipliers and the $27$-bit preadders within the DSP slices through multi-word arithmetic~\cite{HG2006}.
DSP slices have multipliers that are $18 \times 27$-bit wide. Using multi-word arithmetic, we can split $54$-bit operands into multiple $18$-bit operands and operate over them in parallel. 
To perform integer additions, we split the $54$-bit operands into two $27$-bit operands to leverage the $27$-bit preadders in the DSP blocks.
Second, a $54$-bit limb width allows us to make the most of the scarce on-chip memory resources, which includes both Ultra RAM (URAM) and Block-RAM (BRAM).
On U$280$ cards, locations within a URAM block can store $72$-bit wide data and locations within a BRAM block can store $18$-bit wide data. 
Therefore, with $54$-bit (a multiple of $18$) coefficients in the vectors, we can effectively utilize the entire data width of the on-chip memory resources by combining multiple B/URAM blocks to store single/multiple coefficients at a given address.
We discuss a detailed on-chip memory layout later in Section~\ref{sec:Microarchitecture}.

\begin{table}[t]
    \centering
    \caption{Parameter set for FPGA implementation.}
    \label{tab:parameterset}
    \begin{tabular}{p{0.12\columnwidth}p{0.12\columnwidth} p{0.12\columnwidth}p{0.12\columnwidth}p{0.12\columnwidth}p{0.12\columnwidth}}
    \toprule
    $\log q$ & $N$ & $L$ & $\dnum$ & $\fftIter$ & $\lambda$\\
    \midrule
    $54$ & $2^{16}$ & $23$ & $3$ & $4$ & $128$ \\
    \bottomrule
    \end{tabular}
\end{table}

\begin{figure}[t]
  \begin{center}
    \includegraphics[width=1\columnwidth]{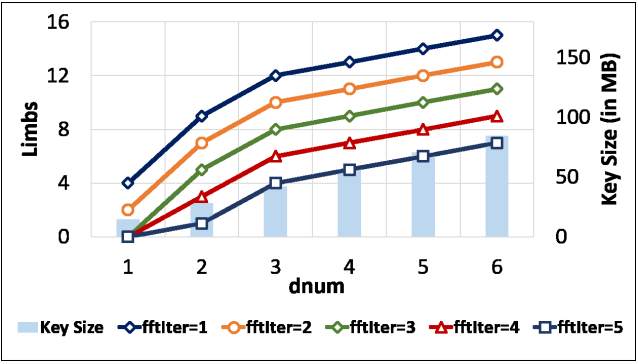}
  \end{center}
  %\vspace{-0.10in}
  \caption{
  Impact of changing the $\dnum$ parameter on the compute levels after bootstrapping and the switching key size. Note that we use the key-compression technique from \cite{fhecompute2021} to halve the size of the keys.}
  \label{fig:dnum}
 \end{figure}

\textbf{Higher-level Parameters.}
The $\dnum$ and $\fftIter$ parameters directly impact a number of factors that determine the final amortized multiplication time per slot, including the bootstrapping runtime as well as the number of compute levels available after bootstrapping.
Figure~\ref{fig:dnum} shows that as we increase the $\dnum$ value, we add more compute levels after bootstrapping, but at the same time we increase the size of $\KeySwitch$ keys, further increasing the compute and on-chip memory requirements. At $\dnum = 3$, we are able to make the best use of the on-chip memory for the corresponding $\KeySwitch$ key size. 

\begin{figure}[t]
  \begin{center}
    \includegraphics[width=1.0\columnwidth]{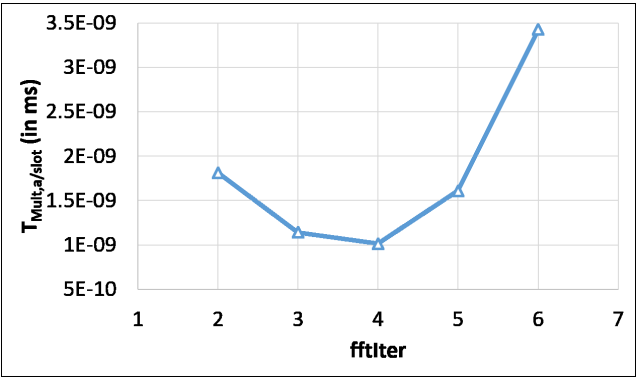}
  \end{center}
  %\vspace{-0.10in}
  \caption{
  Effect of increasing $\fftIter$ parameter on the execution time of bootstrapping. 
  For all benchmarks we set $N = 2^{16}$, $\log(PQ) = 1728$, $\log(q) = 54$, and $\dnum = 3$.}
  \label{fig:fftIter}
 \end{figure}

As mentioned earlier in Section~\ref{subsec:CKKS}, bootstrapping performs an FFT and an inverse FFT as the first and last steps of the algorithm. There is a depth-performance trade-off between the chosen multiplicative depth of the FFT algorithm and the amount of compute required for the bootstrapping operation.
This trade-off is parameterized by $\fftIter$, which is the multiplicative depth of the FFT algorithm.
As the value of $\fftIter$ increases, the multiplicative depth also increases implying fewer compute levels after bootstrapping.
However, as we increase $\fftIter$, the radix of the FFT sub-matrices reduces, thus requiring fewer number of rotations during each multiplication.
Figure~\ref{fig:fftIter} shows how increasing the $\fftIter$ parameter impacts the overall bootstrapping execution time and the number of NTT operations to be computed. The metric used to measure bootstrapping time is the amortized per slot multiplication time in equation~\ref{eq:AmortizedMultTime}.
We pick $\fftIter = 4$ as it generates an optimal balance between the computations (both rotations and NTTs) and the number of compute levels after bootstrapping. This fixes the total depth of our bootstrapping circuit as $L_{\mathsf{Boot}} = 2\cdot \fftIter + 9 = 17$.  

%% file: Sections/FABArchitecture.tex
\section{Overall System Architecture}
\label{sec:Architecture}
\begin{figure*}[t]
 \begin{center}
 \includegraphics[width=6.0in]{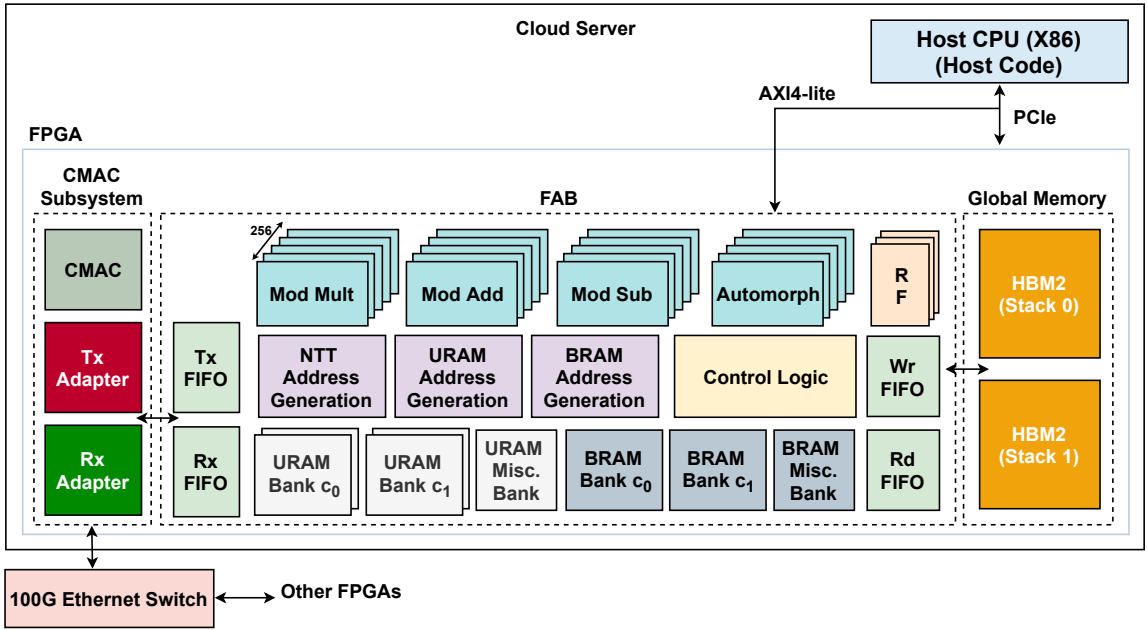}
 \end{center}
 %\vspace{-0.15in}
\caption{An FPGA-based system consisting of a host CPU, FPGA programmable logic, HBM2 memory and CMAC system for FHE-based computing. We map our \fab accelerator to the FPGA programmable logic. The host CPU interacts with the FPGA via PCIe. CMAC subsystem enables interaction between multiple FPGAs via Ethernet Switch.}
\label{fig:FAB}
 % \vspace{-0.1in}
\end{figure*}

In this section, we present the overall architecture (system view) that uses our proposed hardware accelerator. 
As shown in Figure~\ref{fig:FAB}, our overall system consists of four key components: a host CPU (X$86$ in our case) that offloads the RTL design and data to the FPGA, the RTL design that is packaged as a kernel code, global memory on the FPGA comprising of two HBM2 stacks ($4$~GB each), and a $100$G Ethernet (CMAC) subsystem to enable transmit/receive data to/from FPGAs without involving the host.  

On a cloud server, the host CPU is attached to an FPGA accelerator board (Alveo U$280$) via PCIe.
This PCIe interface enables the data transfer between the host and the global memory on the FPGA board.
To enable this data transfer, the host allocates a buffer of the dataset size in the global memory. 
The host code communicates the base address of the buffer with the kernel code using atomic register reads and writes through an AXI4-Lite interface.
The host application also communicates all kernel arguments consisting of the system parameters like prime moduli, the degree of the polynomial modulus $N$, and the pre-computed scalar values (to be stored in the register file) through this interface.
A kernel is started by the host code (written in native C++) using the Xilinx runtime (XRT) API call.
It is worth noting that this XRT API call can be seamlessly replaced by an OpenCL~\cite{munshi2009opencl} API call with trivial modifications to the host code. 
Once the kernel execution starts, no data transfer incurs between the host and the global memory so as to interface all $32$ AXI ports from the HBM to the kernel code. 
The results are transferred back to the host code once the kernel execution completes.

The kernel code instantiates a functional unit consisting of $256$ modular arithmetic and automorph units.
A small register file (RF), $2$~MB in size, stores all the required system parameters and the precomputed scalar values that are received from the host. 
The RF also facilitates temporary storage of up to four polynomials that may be generated as intermediate results.
The kernel has $32$ memory-mapped $256$-bit interfaces that are implemented using AXI4 master interfaces to enable bi-directional data transfers to/from the global memory.
The read (Rd) FIFO and write (Wr) FIFO stream the data from global memory onto the on-chip memory and vice-versa.
The URAM and BRAM resources on the FPGA are organized into various banks to be used as on-chip memory within the kernel code.
All of the URAM memory banks are single-port banks as URAMs do not support dual-port functionality, while the BRAM memory banks are dual-port banks. 
The transmit (Tx) and receive (Rx) FIFO stream the data to and from the CMAC subsystem. 

The Alveo U$280$ FPGA has an integrated IP block for $100$G Ethernet (CMAC) core, providing a high performance, low latency $100$~Gb/s Ethernet port to transfer data between FPGAs that are connected to different hosts.
The CMAC core has an internal clock operating at $322$~MHz and the data interface with the kernel code can be either $256$/$512$~bit wide.
We implement a $512$-bit interface in our kernel code to keep up with $100$~Gbps transfer rates.
This is because, with $512$-bit interface at $300$~MHz, we can theoretically process data at ${\sim} 153$~Gbps, which is faster than the Ethernet IP.
However, with $256$-bit interface, we can process data at ${\sim} 76$~Gbps, which is comparatively slower than the Ethernet IP and will end up dictating the final time it takes to transmit/receive the data to/from other FPGAs. 
Thus, with a $512$-bit interface, it takes ${\sim}11,399$~clock cycles to transmit a single limb of the ciphertext (polynomial of size $0.4$~MB) and ${\sim}546,980$~clock cycles to transmit the entire ciphertext.

%% file: Sections/FABMicroarchitecture.tex
\section{\fab Architecture}
\label{sec:Microarchitecture}
\fab consists of a functional unit, on-chip memory (URAM and BRAM), register file (RF), FIFOs, address generation units, and control logic.
In this section, we discuss the microarchitecture of each of these units.
These units are architected so as to achieve a balanced full-system design.

\subsection{Functional Unit}
\label{subsec:FU}
All operations in FHE break down to integer modular arithmetic i.e., modular addition and modular multiplication operations.
Therefore, each of the $256$~functional units in \fab consists of a modular multiplication, modular addition, modular subtraction, and an automorph unit.
As mentioned in Section~\ref{subsec:Parameter}, we utilize a multi-word arithmetic approach to reduce $54$-bit operations to $27$-bit operations for addition and $18$-bit operations for multiplication.
This facilitates efficient utilization of specialized DSP arithmetic blocks on the FPGA.

For multi-word modular addition and subtraction, we follow Algorithms~$2.7$ and $2.8$ proposed by Hankerson et al.~\cite{HG2006}, respectively.
Both of these algorithms (on line $2$ in Algorithms~$2.7$ and $2.8$) require a correction step for modular reduction, which leads to $54$-bit addition/subtraction operations again.
Subsequently, we modify the correction step in both the algorithms to perform multiple $27$-bit operations instead.
With multi-word arithmetic and all the pipeline registers in place for the DSP blocks, both algorithms perform modular addition and subtraction in $7$~clock cycles.

The modular multiplication is accomplished by first multiplying the operands as integers, and then reducing the result. 
This implies that the modular multiplication is split across two operations i.e., an integer multiplication followed by a modular reduction in a pipelined fashion.
For the integer multiplication, we follow the operand scanning algorithm~\cite{HG2006} (algorithm~$2.9$) that adopts the schoolbook approach to perform multi-word multiplication.
As we split our input $54$-bit operands into three $18$-bit operands, a na\"ive implementation of this algorithm will require $21$~clock cycles to perform a single multiplication.
Given the fact that most FHE workloads have $50\%$ of the operations as integer multiplication, a latency of $21$~clock cycles for a single integer multiplication is too high.
Consequently, we perform loop unrolling on this algorithm and compute various operations in parallel, reducing the multiplication latency to $12$~clock cycles while still adding all the required pipeline registers for DSP multipliers.

\begin{algorithm}[t]
\caption{Modular Reduction in $\mathbb{F}_q$}
\label{algo:MR}
\begin{algorithmic}[1]
\State Modulus $q$, integer $a$, $shifts = 6$ \Comment{a is $(2\log q - 1)$ bits} \label{line:shiftSet}
\State Precompute: $\mathsf{madd}[i-1] = \sum_{j=0}^{5}{i[j] \cdot 2^{\log q + j}\pmod q}$ for $i = 1$ to $2^{shifts}-1$ \Comment{$\mathsf{madd}$ has $\log q$-bit elements} \label{line:MRPrecomp}
\State Set (A[1],A[0]) $\leftarrow$ $a$, $count = 0$, $as_1$ = 0
\While{$count < \log q$}
 \State $(carry,as_1) = A[1] \ll shifts$  \Comment{$carry$ is $shifts$ bit}
 \State $A[1] = as_1 + \mathsf{madd}[carry-1]$  
 \State $count = count + shifts$
\EndWhile
\State $c = A[1] + A[0]$ 
\If {$c > q$}
 \State $c = c - q$
\EndIf \\
\Return {$c$}
\end{algorithmic}
\end{algorithm}
\setlength{\textfloatsep}{3pt}

For modular reduction, we propose a hardware-friendly fast modular reduction algorithm by modifying Will and Ko's reduction technique~\cite{will2014computing}.
As opposed to standard modular reduction approaches like Barrett reduction~\cite{Barrett}, which requires performing multiple expensive multiplication operations, Will and Ko's technique requires only shift and addition operations.
It works by shifting a single bit of the input number, making it a good choice for inputs with smaller bit widths.
For our $(2\log q - 1)$-bit wide number, Will and Ko's technique takes $2\log q$~cycles (for $\log q = 54$ it takes $108$ cycles) to compute a modular reduction.
To reduce this latency while leveraging the simplicity of their approach, we propose Algorithm~\ref{algo:MR} that can instead perform multiple bit shifts at a time requiring only $12$~clock cycles for $\log q = 54$ for the modular reduction operation.
We set the number of shifts to $6$ (line~\ref{line:shiftSet} in \Cref{algo:MR}) in our implementation, but it is worth noting that this algorithm is generic and can work with any number of bit shifts depending on the latency requirement and space constraints.
This algorithm requires precomputing an array $\mathsf{madd}$ (line~\ref{line:MRPrecomp} in \Cref{algo:MR}) having $63$ elements, where each element is $\log q$-bit wide.  
In our case, we need to perform modular reduction w.r.t $32$ different prime moduli, implying that we will need to precompute $32$ such $\mathsf{madd}$ arrays requiring $7$~KB of storage space in total.
However, this precompute is done offline, so there is no compute overhead associated with it.
All the other steps in the proposed algorithm are straightforward and can be performed using inexpensive shift and addition operations.

\begin{figure*}[t]
 \begin{center}
 \includegraphics[width=0.95\textwidth]{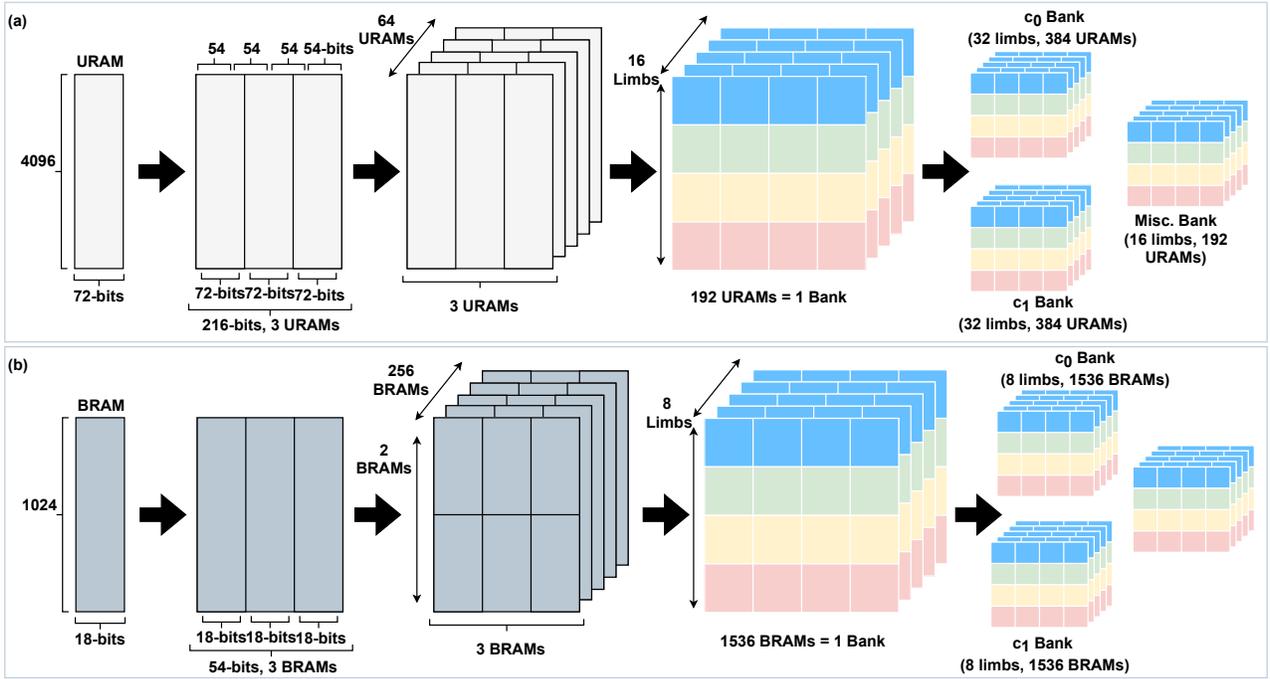}
 \end{center}
  %\vspace{-0.15in}
\caption{On-chip memory configuration; (a) URAM blocks are organized into five different memory banks and (b) BRAM blocks are organized as three different memory banks.}
\label{fig:Onchip}
  %\vspace{-0.2in}
\end{figure*}

The final operation that forms part of the functional unit is $\Automorph$, which performs permutation for the $\Rotate$ operation.
The function of the automorph unit is to read a polynomial from the on-chip memory and store it in the register file in the permuted order as per the given rotation index $k$.
Any original slot indexed by $i$ in a ciphertext maps to the rotated slot through the given automorphism equation:
\begin{equation}
  \mathsf{new\_index}_k(i) = \frac{5^k - 1}{2} + 5 \cdot i \pmod N      
\end{equation}
Due to the limited number of rotation indices (about $60$ different values) being used in bootstrapping, we precompute and store the various powers of $5$ corresponding to each of the rotation index $k$. The division by two is a simple bit-shift, and the reduction modulo $N$ is significantly simplified because $N$ is always a power of two. Thus, reduction modulo $N$ can be achieved by simply performing the \textsf{AND} operation with $N-1$. \\
\noindent \textbf{Summary.} To summarize, the functional units in \fab are highly optimized for hardware to incur less resource overhead. 
They make effective use of high-performance multipliers and adders in DSP blocks to perform low latency modular arithmetic. 
\fab efficiently utilizes these functional units through fine 
grained pipelining and by issuing multiple scalar operations in a single cycle.

\subsection{On-chip Memory}
\label{subsec:onchip}
The Alveo U$280$ accelerator board has single-cycle access URAM and BRAM blocks.
There are $962$~blocks of URAM where each block is $288$~Kb and can be used as single-port memory.
There are $4032$~blocks of BRAM where each block is $18$~Kb and can be used as both single and dual-port memory.
\fab uses a combination of single and dual-port memory banks constructed using URAM and BRAM blocks, providing a total capacity of $43$~MB and a $30$~TB/s internal SRAM bandwidth.

As shown in Figure~\ref{fig:Onchip} (a), each URAM block has data width of $72$-bits and a depth of $4096$.
We combine three such URAM blocks to achieve a data width of $216$-bits.
This allows us to store four $54$-bit coefficients ($216/4 = 54$) at any given address.
Consequently, we need to layout $64$ of these $216$-bit wide URAMs into a single memory bank to enable storage of $256$~coefficients. 
Thus, with every read and write, we can access $256$~coefficients in the same cycle, aligning with the number of functional units in the design.
With this layout, a single memory bank consists of $64 \times 3 = 192$ URAMs and can store $16$~polynomials (${\sim}7.08$~MB). 
We organize the available URAM blocks into five such banks that are divided as follows: 
The first two banks ($\c_0$ bank-$1$ and $2$) store $32$~limbs ($24$ original and $8$ extension) of the $\c_0$ ring element of the ciphertext.
The next two banks ($\c_1$ bank-$1$ and $2$) store $32$~limbs ($24$ original and $8$ extension) of the $\c_1$ ring element of the ciphertext.
The fifth bank can store $16$~polynomials. 
We call the fifth bank the "miscellaneous" bank as it is used to store multiple data items such as twiddle factors, $\KeySwitch$ keys, and plaintext vectors that are read in from the main memory.

As shown in Figure~\ref{fig:Onchip} (b), BRAM blocks are organized as $54$-bit wide memory banks by combining three $18$-bit wide BRAMs.
Since each address can store only a single $54$-bit coefficient, we need $256$~BRAM blocks to store $256$~coefficients.
In addition, the depth of each BRAM block is only $1024$; therefore, we stack two BRAM blocks to get a depth of $2048$, thus enabling storage of $8$~polynomials in a single BRAM bank.
Similar to URAM bank organization, we organize BRAM blocks into multiple banks.
We have three BRAM banks in total, where two banks consists of $1536$ BRAMs each and can store $8$~polynomials and thus, are ideal to store the extension limbs.
While the third bank consists of $768$ BRAMs and can store $4$~polynomials. 
We again call this third bank the "miscellaneous" bank and use it to store temporary data from main memory during various operations.

\noindent \textbf{Summary.} To summarize, \fab architecture efficiently utilizes the available U/BRAM blocks on the FPGA as on-chip memory.
Mapping the polynomial data bit-width to that of the U/BRAM blocks data width enables storage of up to $43$~MB of on-chip data.
\fab overcomes the limited main memory bandwidth issue by utilizing a combination of single and dual-port memory banks that complement the operational needs of the underlying FHE operations, resulting into a more balanced FPGA design. 

\subsection{Register File}
\label{subsec:RF}
Our design consists of multiple register files (RFs).
The total capacity of all the register files is $2$~MB.
The register files are spread across the design and are used by functional, address generation and control units. 
Each RF has multiple read/write ports with same cycle access latency.
About one-fourth of the register file is used to store pre-computed values and system parameters, which are written by the host CPU through atomic writes before launching the kernel code execution.
The remaining RFs are used to store up to four intermediate polynomials that are generated as part of $\Rotate$ or $\Mult$ operations.

\subsection{FIFOs}
\label{subsec:FIFO}
We instantiate $32$ synchronous Write (Wr) and Read (Rd) FIFOs (supporting $32$ AXI ports on the HBM-side) to stream the data between the main memory and the on-chip memory.
These FIFOs are composed of the distributed RAM available on the FPGA board.
The data width of each FIFO is equal to the data width supported by each AXI port i.e., $256$-bits.
The depth of the Wr FIFO is $128$ to support an HBM burst length of $128$.
The depth of the Rd FIFO is $512$ to support up to four outstanding reads.
Rd FIFO is driven by the memory-side clock domain having a clock frequency of $450$~MHz while the Wr FIFO is driven by the kernel-side clock domain having a clock frequency of $300$~MHz.
We also instantiate a Transmit (Tx) and Receive (Rx) FIFO to stream the data between the CMAC subsystem and the on-chip memory.
These are also synchronous FIFOs having a $512$-bit data interface.

\begin{figure*}[t]
 \begin{center}
 \includegraphics[width=0.93\textwidth]{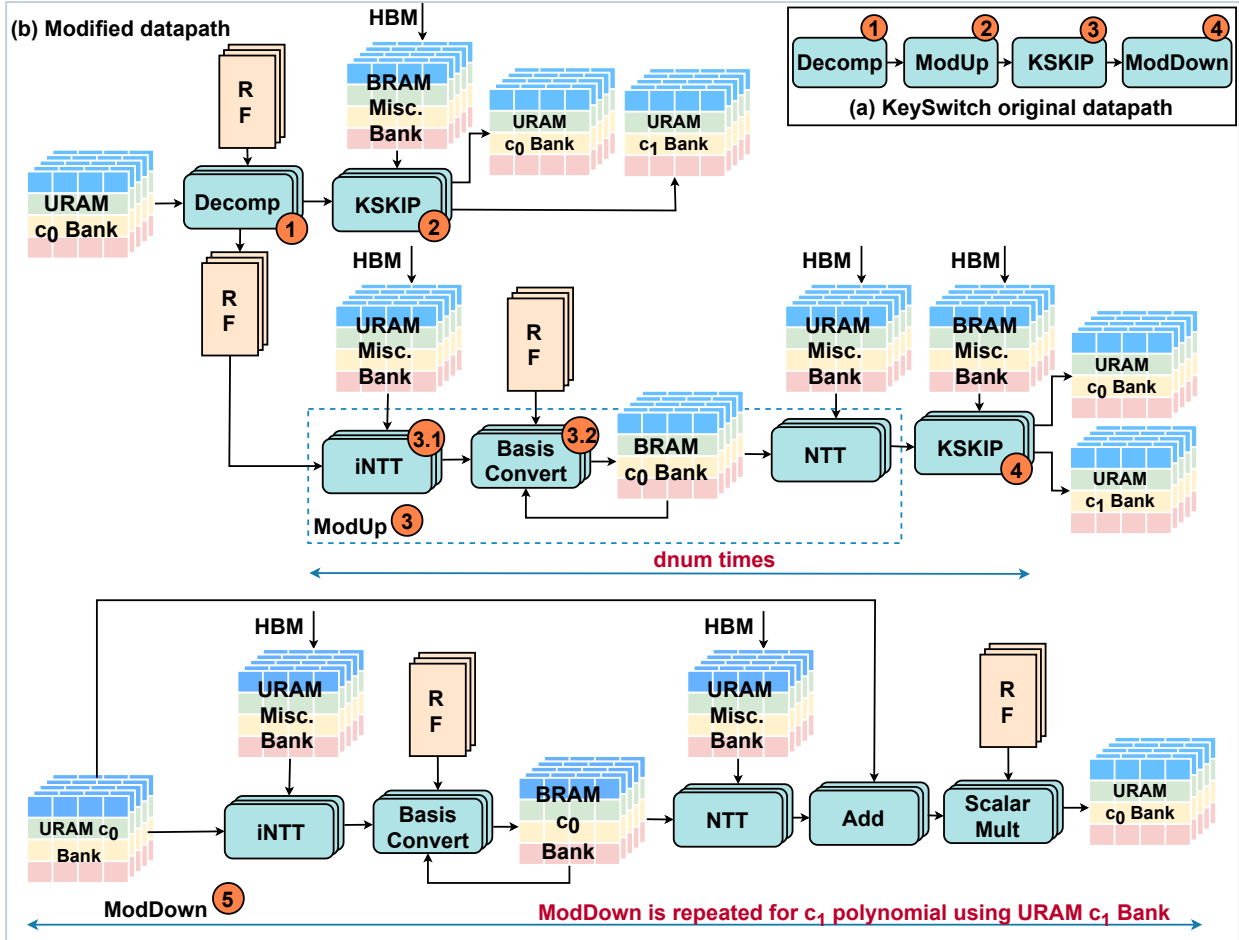}
 \end{center}
  %\vspace{-0.25in} 
\caption{(a) Original datapath. (b) Modified datapath for efficient on-chip memory utilization for the four main operations ($\Decomp$, $\ModUp$, $\KSKInProd$, and $\ModDown$) within the $\KeySwitch$ Operation. Our approach avoids reads/writes of the ciphertext limbs to the main memory, which lowers the latency of FHE-based computing.} 
\label{fig:KS}
  %\vspace{-0.2in}
\end{figure*}

\subsection{NTT/iNTT datapath}
\label{subsec:NTT}
Our NTT datapath uses a unified Cooley-Tukey algorithm \cite{norton1987parallelization} for both NTT and inverse-NTT (iNTT).
Using a unified NTT algorithm provides the convenience of leveraging the same data mapping logic for both NTT and iNTT.
The $256$ modular addition, subtraction and multiplication units operate in parallel as radix-$2$ butterfly units, processing $512$ coefficients of a polynomial at once. 
This allows us to perform $\log N$ stages in approximately $\log N \cdot \frac{N}{512}$~cycles instead of $\log N \cdot \frac{N}{2}$~cycles.
The NTT address generation unit (shown in Figure~\ref{fig:FAB}) takes care of uniquely mapping the data within each stage of the NTT/iNTT using a sub-unit, i.e. a data mapping unit.
Furthermore, a twiddle factor mapping sub-unit within the NTT address generation unit takes care of reading the required twiddle factors for an NTT stage from the URAM miscellaneous bank.  
Both of these sub-units leverage the data and stage counters to generate the addresses on-the-fly using inexpensive shift, and AND operations. 
Thus, we efficiently leverage pipelining and parallelism while computing NTT/iNTT by spreading the computations over the functional units, a data mapping unit, and a twiddle factor mapping unit. 
It is worth noting that we do not take into account the latency of the bit-reversal operation here as bit-reversal is carried out along with automorph/multiplication operation that is performed just before NTT/iNTT. 

\subsection{Key Switch datapath}
\label{subsec:keyswitch}
A $\KeySwitch$ operation comprises of four sub-operations, i.e., $\Decomp$, $\ModUp$, $\KSKInProd$, and $\ModDown$.
With limited on-chip memory, these operations require smart operation scheduling to efficiently utilize the on-chip memory.
This is because $\KeySwitch$ not only needs to operate on the extension limbs (the factors of $P$) but it also needs to perform inner product with the $\KeySwitch$ keys that are almost three times the size of our ciphertext.
\textbf{Below we present a detailed description on how we schedule and reorganize the sub-operations in $\KeySwitch$ to manage ${\sim}112$~MB of data ($84$~MB keys and $28$~MB ciphertext) within the available $43$~MB on-chip memory without the need of writing any resultant limbs back to the main memory.}

The $\Decomp$ sub-operation splits the limbs in $\a_\m$ ring element (ciphertext has two ring elements $\a_\m$ and $\b_\m$) into $\dnum$ digits. 
Each of these digits is then passed into $\ModUp$. 
$\ModUp$ outputs $L+1+\alpha$ limbs; after all $\ModUp$ operations are complete we have $\dnum \cdot (L+1+\alpha)$ limbs in total.
In our case, $\dnum = 3$ and $\alpha = 8$. 
This results in $3\cdot 32 = 96$ total limbs after all instances of the $\ModUp$ operation have finished.
Therefore, to manage all $96$~limbs in on-chip memory (without having to write any resultant limb back to main memory), we modify the $\KeySwitch$ datapath to reorganize the execution of sub-operations in $\KeySwitch$. 
Na\"ively, one would execute the sub-operations one after another following the original datapath as shown in Figure~\ref{fig:KS} (a).
However, there are challenges associated with this approach that \fab addresses through the modified datapath (refer Figure~\ref{fig:KS} (b)) and smart operation scheduling.

\noindent \textbf{Modified datapath:}
We modify the original $\KeySwitch$ datapath so as to split the $\KSKInProd$ step into two steps. 
Rather than perform the $\KSKInProd$ step all at once, we "greedily" make progress on the inner product by performing the multiplications and additions as soon as the operands are in memory, which saves DRAM transfers.\\ \indent
In more detail, the $\KSKInProd$ step is an inner product between the input polynomial and the switching key in the raised basis. 
In the original $\KeySwitch$ data path, the full $\ModUp$ step is performed to extend the RNS basis of the input polynomial, then the full $\KSKInProd$ is performed. 
There is not enough space in the on-chip memory to hold all the extended polynomials, and so the original datapath first reads in the input limbs in evaluation representation to perform the $\Decomp$ step, then writes the limbs out in coefficient representation after the $\ModUp$ step. 
These limbs are then read back into the chip memory to perform the $\KSKInProd$ step. \\ \indent
This optimization identifies that $\alpha$ limbs in each $\ModUp$ operation do not change, so we can use these limbs to immediately begin the $\KSKInProd$ step once the $\Decomp$ step is finished. 
More specifically, the $\Decomp$ step takes these $L$ limbs and splits them into $\beta \leq \dnum$ blocks of $\alpha$ limbs each. 
These $\alpha$ limbs then take two paths. 
First, these $\alpha$ limbs are used to begin the $\KSKInProd$, and the intermediate sum is written out to URAM. 
The second path these $\alpha$ limbs take is as input to the basis conversion step so that the extension limbs can be generated. 
Once these extension limbs are generated, they are used to complete the $\KSKInProd$ step.
\textbf{Thus, the modified datapath not only reduces the number of NTT computations (the most expensive subroutine in $\KeySwitch$ operation) but also helps alleviate the memory bandwidth bottleneck by reducing memory traffic.}

\noindent \textbf{Smart operation scheduling:}
Smart operation scheduling takes advantage of a straightforward optimization in the RNS extension equation in \Cref{eq:RNSExtension}. 
Naively, for each new limb $p_j$ that must be generated, we must perform $2\ell$ multiplications. 
However, we can observe that one of the multiplications (specifically the $x_i \cdot \tilde{Q}_i$ product) is not dependent on the output modulus. 
Therefore, we can compute the $\ell$ inner multiplications of the form $x_i \cdot \tilde{Q}_i \pmod{q_i}$ and then reuse these products when computing all of the extension limbs $p_j$. 
This reduces the number of modular multiplications by a factor of two. \\ \indent
Smart operation scheduling also complements the modified datapath by generating an entire block of extension limbs at once for the given the original $\alpha$ limbs.
As soon as all of the extension limbs for one block of $\alpha$ limbs is computed, we perform an NTT followed by $\KSKInProd$ on these limbs before generating the next block of extension limbs. 
This process is repeated $\dnum$ times to generate all $\dnum \cdot (L + 1 + \alpha)$ limbs.
Note that we do not need to fetch all the $\KeySwitch$ keys at once; we only need to fetch the block of the key corresponding to the $\alpha$ block that has just been extended.
Moreover, all extension limbs are always stored in $\c_0$ BRAM bank because the dual-port BRAMs supplement $\BasisConvert$ operation by allowing data reads and writes at the same time to perform inner products in \emph{limb-wise} fashion.
Through these methods, we perform the entire $\KSKInProd$ using a small BRAM miscellaneous bank (that can store only $4$ limbs), while hiding the key read latency (which is about $300$~clock cycles) from the main memory behind the computations. 
The $\ModDown$ (step-$5$ in Figure~\ref{fig:KS} (b)) operation follows a similar operation scheduling as $\ModUp$ operation.
\textbf{Thus, through smart scheduling we enable high data reuse, exploit inherent \emph{limb-wise} parallelism and maintain a uniform address generation logic by avoiding to switch between \emph{limb-wise} and \emph{slot-wise} accesses, and further reduce the main memory traffic by not writing/reading any resultant ciphertext limbs to the main memory.}

We would like to mention that the split of the $\KSKInProd$ into two steps does not change the underlying $\KeySwitch$ algorithm, only the order in which the steps are performed changes. 
The resulting noise from the $\KeySwitch$ algorithm is identical with or without this reordering.
Although our $\KeySwitch$ datapath optimization is specific to Alveo $U280$ FPGA, it can be ported to smaller FPGAs as long as one can accommodate at least one limb of the key and the ciphertext polynomial in on-chip memory. 
If not, then some additional fine grain optimizations will have to be developed such as changing the $\BasisConvert$ operation to perform all operations \emph{slot-wise} instead of \emph{limb-wise}. 
This is to efficiently manage reading the slots from within a key limb so that the problem does not become memory bound.

%% file: Sections/Evaluation.tex
\section{Evaluation}
\label{sec:Evaluation}
\vspace{-0.05in}

\subsection{Setup}
\label{subsec:Setup}
We designed \fab in Verilog $2001$ and synthesized it using Xilinx Vivado $2020.2$ to operate at $300$~MHz frequency.
The host CPU code is written in C++.
\fab RTL code is packaged into kernel code using the Xilinx Vitis $2020.2$ development platform.
The kernel code is compiled and linked into an FPGA executable (.xclbin binary file) by the Vitis compiler.
In the cloud environment, this binary file is mapped to a Xilinx Alveo U$280$ FPGA and its execution is initiated by the host CPU via host code. 
This accelerator FPGA card is built on the Xilinx $16$nm UltraScale architecture and offers $8$~GB of HBM$2$ with up to $460$~GB/s bandwidth.

\subsection{Resource Utilization}
\label{subsec:HR}
Table~\ref{tab:FABResource} provides the hardware resource utilization of the various components of \fab.
Overall, \fab requires ${\sim}899$K LUTs, and the functional units represent the largest share (${\sim}37\%$) of these LUTs.
The remaining utilization is among the control unit, the various address generation units and the FIFOs.
Out of the $2,073$K flip flops (FFs) used by \fab, most of them are utilized by the distributed register file, control logic and the functional unit.
The entire $56.7\%$ of the DSP utilization is for the modular arithmetic operations within the functional units.
As mentioned earlier in Section~\ref{subsec:onchip}, \fab makes use of almost the entire URAM and BRAM blocks on FPGA; we observe a $95.24\%$ BRAM utilization and $99.8\%$ URAM utilization.

\begin{table}[t]
    \centering
    \caption{FAB Hardware Resource Utilization}
    \label{tab:FABResource}
    \begin{tabular}{cccc}
    \toprule
    \textbf{Resource} & \textbf{Available} & \textbf{Utilized} & \textbf{\% Utilization} \\
    \midrule
    LUTs & $1,304$K & $899,232$ & $68.96$  \\
    FFs & $2,607$K & $2,073$K & $79.54$  \\
    DSP & $9,024$ & $5,120$ & $56.70$  \\
    BRAM & $4,032$& $3,840$ & $95.24$  \\
    URAM & $962$ & $960$ & $99.80$  \\
    \bottomrule
    \end{tabular}
\end{table}

\begin{table}[t]
    \centering
    \caption{Comparison of modular multiplier count, register file size and on-chip memory size in different designs.}
    \label{tab:ResourceComp}
    \begin{tabular}{p{0.13\columnwidth} p{0.18\columnwidth} p{0.15\columnwidth} p{0.15\columnwidth} p{0.15\columnwidth}}
    \toprule
    \textbf{Work} & 
    \textbf{Parameters (N, $\log q$)} & 
    \textbf{Modular multipliers} & 
    \textbf{Register file (MB)} & 
    \textbf{On-chip memory (MB)} \\
    \midrule
    F1~\cite{samardzic2021f1} & $2^{14}$, $32$ & $18432$ & $8$ & $64$ \\
    BTS~\cite{kim2021bts} & $2^{17}$, $50$ & $8192$ & $22$ & $512$ \\
    \fab & $2^{16}$, $54$ & $256$ & $2$ & $43$ \\
    \bottomrule
    \end{tabular}
\end{table}

\subsection{Comparison of Basic FHE Operations}
\label{subsec:OperationsLatency}
Table~\ref{tab:OpLatency} presents the execution time (in ms) for basic operations in the CKKS FHE scheme and compares the performance with the existing GPU implementation~\cite{jung2021over}.
The GPU numbers used in the table are the most optimized performance numbers reported by the authors for the parameter set $N=2^{16}$, $\log Q = 1693$, and $100$b security.
We compare against these numbers as the parameter set is closest to our parameters.
\fab achieves an average $2.4\times$ speedup when compared to GPU in absolute execution time for these basic primitives.
For completeness, in Table~\ref{tab:OpThroughput}, we compare the performance of $\NTT$ and $\Mult$ with an existing state-of-the-art FPGA implementation (HEAX~\cite{riazi2020heax}).
For a fair comparison, we use the same parameter set ($N = 2^{14}$ and $\log Q = 438$) as used in HEAX.
\fab achieves an average $3\times$ higher throughput (in operations per second) when compared to HEAX. 
The performance gain in \fab is largely due to low latency modular arithmetic modules within functional units, fine-grained pipelined usage of the functional units, highly optimized $\NTT$ datapath, and the modified $\KeySwitch$ datapath. \\ \indent
We do not compare \fab against ASIC implementations of CKKS such as F$1$~\cite{samardzic2021f1} that do not support parameters large enough for fully-packed bootstrapping. 
Also note that several works~\cite{BMTH20, kim2021bts} do not give low-level benchmarks for the individual homomorphic operation. 
For these works, we compare against the amortized multiplication time in \Cref{tab:BLatency}.

\begin{table}[t]
    \centering
    \caption{Execution time (in ms) for performing basic CKKS FHE operations and speedup achieved using \fab.}
    \label{tab:OpLatency}
    \begin{tabular}{p{0.25\columnwidth} p{0.12\columnwidth} p{0.15\columnwidth} p{0.30\columnwidth}}
    \toprule
    \textbf{Operation} & 
    \textbf{\fab} & 
    \textbf{GPU} & %~\cite{jung2021over} & 
    \textbf{Speedup vs GPU} \\
    \midrule
    $\Add$ & $0.04$ & $0.16$ & $3.85\times$ \\
    $\Mult$ & $1.71$ & $2.96$ & $1.73\times$ \\
    $\Rescale$ & $0.19$ & $0.49$ & $2.62\times$ \\
    $\Rotate$  & $1.57$ & $2.55$ & $1.62\times$ \\
    \bottomrule
    \end{tabular}
\end{table}

\begin{table}[t]
    \centering
    \caption{Throughput (in operations per second) comparison for basic operations with HEAX. Throughput numbers reported here are for $N = 2^{14}$ and $\log Q = 438$.}
    \label{tab:OpThroughput}
    \begin{tabular}{p{0.20\columnwidth} p{0.14\columnwidth} p{0.14\columnwidth} p{0.33\columnwidth}}
    \toprule
    \textbf{Operation} & \textbf{\fab} & \textbf{HEAX} & \textbf{Speedup vs HEAX} \\
    \midrule
    $\NTT$ & $167$K & $42$K & $3.97\times$ \\
    $\Mult$ & $5.7$K & $2.6$K & $2.12\times$ \\
    \bottomrule
    \end{tabular}
\end{table}

\subsection{Bootstrapping Latency}
\label{subsec:BootstrappingLatency}
As described earlier, the bootstrapping operation is the key bottleneck for performing unbounded FHE computations.
In this section, we compare the bootstrapping latency of \fab with the existing state-of-the-art CPU, GPU, and ASIC implementations.
Throughout this section, we refer to these implementations as Lattigo~\cite{BMTH20} (CPU implementation), GPU-$1$ for $97$-bit security and GPU-$2$ for $173$-bit security (GPU implementations of~\cite{jung2021over}), F$1$~\cite{samardzic2021f1}, and BTS-$2$~\cite{kim2021bts} (ASIC implementations).  
Following the bootstrapping performance metric from these existing works, we compare against these works using the amortized per slot multiplication time $T_{\Mult,a/slot}$ defined in equation~\ref{eq:AmortizedMultTime}.

As seen in Table~\ref{tab:BLatency}, \fab outperforms CPU, GPU-$1$, GPU-$2$ and F$1$ implementations in absolute time. 
This is despite the lower operating frequency of \fab, and when comparing the clock cycles against prior works \fab compares even more favorably. 
\fab exhibits a better performance due to improved arithmetic intensity by overcoming memory bandwidth bottleneck. 
However, \fab is about $9\times$ (absolute time) and $4\times$ (clock cycles) slower than the best case numbers reported by BTS (i.e., for BTS-$2$).
This difference in performance is mainly because of the large on-chip memory and a large number of modular multipliers used for computation in BTS when compared to \fab. 

In Table~\ref{tab:ResourceComp} we highlight the difference in the resources utilized by the ASIC implementations and \fab to achieve the aforementioned performance.
We perform this comparison in terms of the number of modular multipliers (MMs) used, the size of the register file and the on-chip memory.
When compared to BTS, \fab requires $32\times$ less MMs, $11\times$ smaller register file and $12\times$ smaller on-chip memory.
Moreover, the BTS design uses an ASAP$7$ technology node leading to an expensive design (in terms of fabrication cost and engineering design team cost).
In contrast, we use FPGAs that are available today and our solution can be readily deployed in today's cloud systems.
If \fab were to implement $8192$ MMs and a $512$~MB on-chip memory, \fab will be at least $3\times$ faster than BTS owing to its optimization to the bootstrapping algorithm and the microarchitecture implementation. 

\begin{table}[t]
    \centering
    \caption{Speedup achieved using \fab when performing bootstrapping operations. Slots define the number of packed slots in the ciphertext while bootstrapping. Bootstrapping time is computed in $T_{mult,a/slot}$.}
    \label{tab:BLatency}
    \begin{tabular}{p{0.12\columnwidth} p{0.11\columnwidth} p{0.09\columnwidth} p{0.12\columnwidth} p{0.13\columnwidth} p{0.15\columnwidth}}
    \toprule
    \textbf{Work} & 
    \textbf{Freq. (in GHz)} &  
    \textbf{Slots} & 
    \textbf{Time (in $\mu s$)} & 
    \textbf{Speedup achieved using \fab (Time)} & 
    \textbf{Speedup achieved using \fab (Cycles)} \\
    \midrule
    Lattigo & $3.5$ & $2^{15}$ & $101.78$ & $213\times$ & $2485\times$ \\
    GPU-$1$ & $1.2$ & $2^{15}$ & $0.740$ & $1.55\times$ & $6.35\times$ \\
    GPU-$2$ & $1.2$ & $2^{16}$ & $0.716$ & $1.50\times$ & $6.14\times$ \\
    F$1$  &  $1$ & $1$ & $254.46$ & $533\times$ & $1775\times$ \\
    BTS-$2$ & $1.2$ & $2^{16}$ & $0.0455$ & $0.09\times$ & $0.38\times$ \\
    \fab & $0.3$ & $2^{15}$ & $0.477$ & - & - \\
    \bottomrule
    \end{tabular}
\end{table}

\subsection{Logistic Regression (LR) Training}
\label{subsec:LR}
In this section, we evaluate the use of \fab to perform LR model training for binary classification over a subset of MNIST data~\cite{deng2012mnist} labeled $3$ and $8$.
This is the task considered in the HELR work~\cite{han2018efficient}, and it is the same task used to benchmark all works we compare against.

This subset of the dataset contains $11,982$ training samples where each sample has $196$~features.
The LR model is trained for $30$~iterations with $1024$ encrypted images in a mini-batch.
We adopt the sequence of operations proposed by Han et al.~\cite{han2018efficient} for efficient logistic regression training on encrypted data. 
Using this algorithm, LR training for $30$~iterations has an evaluation depth of $150$, and thus, it requires us to perform a bootstrapping operation after every iteration.

We perform training using sparsely-packed ciphertexts (using $256$ slots only). 
This is largely to perform a fair comparison with existing works (GPU-$2$ and BTS-$2$) that have only considered sparsely-packed ciphertexts ($256$ slots) for LR training. 
This is because $256$ slots is optimal for the specific benchmark task, but our LR training implementation can easily scale to larger applications (i.e. applications that require fully-packed bootstrapping). 

To evaluate LR training, we present two different FPGA designs here; 1) \fab-$1$: a single-FPGA design and 2) \fab-$2$: a multi-FPGA design that utilizes eight FPGAs. 
Note that \fab-$2$ will incur eight times the resource utilization as \fab-$1$.
We follow the data partitioning and packing technique proposed in Han et al.~\cite{han2018efficient} to pack the data efficiently into ciphertexts.
Our \fab-$1$ design is a straightforward mapping of \fab onto an Alveo U$280$ board with all the ciphertexts and $\KeySwitch$ keys (${\sim}6.65$~GB data) offloaded to the HBM$2$ of the FPGA. 
For the \fab-$2$ design, we instantiate \fab on all eight FPGA boards in the cloud environment, where each FPGA is connected to a host CPU.
We form FPGA pairs, and in each pair, we designate a primary FPGA and a secondary FPGA to enable point-to-point communication between the FPGAs.
In addition, one of the eight FPGAs acts as a master FPGA that can broadcast a ciphertext to the entire pool of FPGAs. 
We limit the impact of network communication on the performance of our \fab-$2$ design by using direct network communication between the FPGAs instead of involving the host CPUs.
All the host CPUs launch the required kernel code on their respective FPGAs in parallel, which allows multiple ciphertexts to be processed in parallel.
Since the ciphertexts are sparsely-packed, each FPGA needs to compute on $128$~ciphertexts.
Communication between the FPGAs needs to happen only twice during the execution of an entire LR iteration. 
This communication overhead is about $12$ms per LR iteration.

Table~\ref{tab:LRLatency} presents the average training time per LR iteration.
In terms of absolute execution times, \fab-$2$ is $456\times$ and $9.5\times$ faster than existing state-of-the-art CPU and GPU implementations (for parameter set having $N=2^{17}$ and $\log Q=2395$).
\fab-$2$ is even faster when we compare the speedup in terms of clock cycles.
When comparing to ASIC proposals, \fab-$2$ outperforms F$1$ by $12\times$ 
while achieving a competitive performance when compared to BTS-$2$.  
Compared to \fab-$1$, \fab-$2$ (using eight FPGAs) does not observe a corresponding $8\times$ speedup as the amount of parallelism that can be extracted is limited by the bootstrapping runtime (following Amdahl's law). 
This is because \fab is designed to perform bootstrapping on a single FPGA, so the performance increase is due to parallelizing the other operations within a logistic regression iteration. 
In principle, it is possible to improve the performance of \fab-$2$ by scaling bootstrapping operation to multiple FPGAs by exploiting instruction level parallelism.
However, this requires distributing the operations (on a single ciphertext) to multiple FPGAs while minimizing data hazards and managing the large communication overhead. 
This is part of our future work.

\begin{table}[t]
    \centering
    \caption{Performance comparison for LR training when using sparsely-packed ciphertexts~\cite{kim2021bts}. Time reported here is average training time per iteration.}
    \label{tab:LRLatency}
    \begin{tabular}{p{0.12\columnwidth} p{0.1\columnwidth} p{0.3\columnwidth} p{0.3\columnwidth}}
    \toprule
    \textbf{Work} & 
    \textbf{Time (in sec)} & 
    \textbf{Speedup achieved using \fab-$2$ (Time)} & 
    \textbf{Speedup achieved using \fab-$2$ (Cycles)} \\
    \midrule
    Lattigo  & $37.05$  & $456\times$ & $5318\times$ \\
    GPU-2    & $0.775$  & $9.5\times$ & $39\times$  \\
    F$1$     & $1.024$  & $12\times$  & $41\times$  \\
    BTS-$2$  & $0.028$  & $0.3\times$ & $1.4\times$  \\
    \fab-$1$ & $0.103$  & $1.3\times$ & $1.3\times$ \\
    \fab-$2$ & $0.081$  & -           & - \\
    \bottomrule
    \end{tabular}
\end{table}

\noindent \textbf{Comparison with Leveled FHE Approach:}
Here we briefly compare our bootstrapping-based FHE approach against the leveled FHE approach. 
The leveled FHE approach avoids bootstrapping by simply having the cloud host send a ciphertext with no remaining compute levels back to the client, who holds the decryption key. 
The client then decrypts the ciphertext and re-encrypts the resulting message into a new ciphertext with some fixed number of compute levels. 
This new ciphertext is sent back to the cloud host to proceed with the homomorphic computation.\\ \indent
There are several drawbacks to this approach. 
The most immediate is information leakage, which could make this approach completely unusable in some applications. 
This leakage is due to the intermediate values of the computation being revealed to the client; in contrast, only the final output value is revealed to the client when bootstrapping is run on the cloud machine. 
In order to defend against this leakage, the cloud host will need to add a $\lambda$-bit mask to the plaintext in order to achieve $\lambda$ bits of security. 
This is because the CKKS plaintext space is \emph{not} finite\footnote{This is in contrast to FHE scheme such as BGV~\cite{BGV12} and BFV~\cite{Brak12, FV12}, which have finite plaintext spaces. For these schemes, a uniformly random mask can be added to hide the message without requiring any additional bits of precision.}.
In order to maintain correctness, this requires the CKKS plaintext to support an additional $\lambda$ bits beyond the original plaintext value.
The minimum $\lambda$ that could be justified is $\lambda=40$, which would require a significant increase in the CKKS parameters and result in a significant increase in the LR training time. \\ \indent
Even if we ignore the information leakage, \fab still outperforms the leveled FHE approach. 
A single iteration of logistic regression consumes $5$ compute levels and requires bootstrapping at the end of each iteration. 
Using \fab, one iteration takes $0.103$ seconds, which includes the bootstrapping time. 
For the leveled FHE approach, the encryption on the client side itself (using the sub-routines in the SEAL library) takes $0.162$ seconds with a $2.8$ GHz CPU.
Adding the time for operations in the cloud and the time for client-cloud communication further increases the time per LR iteration.

%% file: Sections/RelatedWork.tex
\section{Related Work}
\label{sec:RelatedWork}
\noindent \textbf{CPU/GPU-based Acceleration:}
Many software libraries such as 
SEAL~\cite{sealcrypto}, HELib~\cite{HElib,HS2014}, PALISADE~\cite{PL2019}, Lattigo~\cite{LT2019}, and HEAAN~\cite{KH2018} 
implement the CKKS scheme on CPU. Despite the efforts of these libraries, a pure-CPU implementation of FHE remains impractical.
A number of works~\cite{kim2020,zhai,goeyGPU} focus on accelerating just the NTT computation on a GPU.
The current state-of-the-art for pure-CPU/GPU implementation of the full suite of FHE (including basic operations and bootstrapping) is the work of Jung et al.\cite{jung2021over} which proposed the first GPU-accelerated implementation of the CKKS scheme. \\
\noindent \textbf{ASIC-based Acceleration:}
Samardzic et al.~\cite{samardzic2021f1} present the F$1$ hardware accelerator architecture for FHE computations. 
Although the accelerator supports multiple FHE schemes including BGV~\cite{gentry2012ring}, CKKS, and GSW~\cite{hiromasa2016packing}, it only supports operations on smaller parameter sets.  
To support multi-slot bootstrapping and the larger applications that require multi-slot bootstrapping (e.g. real-time machine learning), larger parameters are required, making the F$1$ chip incompatible with these applications.
Kim et al.~\cite{kim2021bts} proposed the BTS accelerator specifically tailored to efficiently support CKKS bootstrapping. 
BTS employs a massive number of processing elements to exploit parallelism in various homomorphic operations and makes use of large on-chip memory with a large register file. Despite these resources, they do not modify the underlying algorithm to fully optimize for the memory bandwidth. This results in  unrealized performance when compared to the resources they use. \\
\noindent \textbf{FPGA-based Acceleration:}
FPGAs are a more suitable candidate for high-performance and low-power secure computation.
HEAX~\cite{riazi2020heax} is an FPGA-based accelerator that accelerates only CKKS encrypted multiplication.
Other operations are deferred to a host processor.
Moreover, this FPGA implementation supports only smaller parameter sets that allow computation up to multiplicative depth $10$, which is not sufficient for bootstrapping or applications such as logistic regression training.
Given these limitations, it is unclear how to implement a full-scale CKKS FHE workload and an end-to-end application on an FPGA.
In our work, we implement the full-scale CKKS FHE bootstrapping, and we do so with secure parameter sets that support end-to-end applications. 
We evaluate a full system for secure logistic regression on an FPGA. \\
\noindent \textbf{Applicability to Other Schemes:}
Although \fab implements the CKKS scheme-specific bootstrapping, our implementations of the basic operations such as $\Add$, $\Mult$, and $\Rotate$ that are common across schemes can be used for the BGV~\cite{BGV12} and B/FV~\cite{Brak12,FV12} schemes.
FHE schemes like TFHE~\cite{chillotti2020tfhe} and FHEW~\cite{ducas2015fhew} evaluate Boolean gates on encrypted data while incurring several GB memory footprint for encrypted keys~\cite{gupta2022memfhe}.
Even for these schemes optimizations similar to $\KeySwitch$ datapath and smart operation scheduling are very relevant.
However, the specific steps within the $\KeySwitch$ operation for all of these other schemes differ from those in the CKKS scheme, and so a thorough analysis is required to determine the exact operation scheduling so as to develop a balanced FPGA design.  

%% file: Sections/Conclusion.tex
\section{Conclusion}
\label{sec:Conclusion}
We propose \fab, an FPGA-based accelerator for bootstrappable FHE. 
\fab leverages a combination of algorithmic and architectural optimizations to overcome the memory bandwidth bottleneck and to perform the first ever fully-packed bootstrapping on FPGA for a practical parameter set.
Through smart operation scheduling and memory management techniques, \fab efficiently utilizes the limited compute and memory resources on the FPGA to deliver $456\times$ and $6.5\times$ better performance than CPU and GPU, respectively, for LR training application.
More importantly, for the same LR training application \fab uses only currently-existing hardware while delivering performance comparable to the specialized and costly ASIC designs for bootstrappable FHE. 
\fab is also immediately accessible to the general public as all the resources required to support \fab exist in public commercial cloud environments.